\documentclass[aps,prl,showkeys,floatfix,amsmath,amssymb,reprint]{revtex4-1}
\usepackage{graphicx} 
\usepackage{bm}
\usepackage{color,mathtools,bbm} 
\usepackage{wasysym}    
\usepackage{hyperref}  
\usepackage{lineno}
\usepackage{verbatim}
\usepackage{longtable}
\usepackage{footnotehyper}
\usepackage{multirow} 
\usepackage{mciteplus}
\usepackage{array}
\usepackage{hypernat}
\usepackage{float} 
\usepackage{wrapfig} 
\usepackage{adjustbox}
\usepackage{braket}
\usepackage{amsmath}
\usepackage{fancyvrb}
\usepackage{listings}

\usepackage{enumitem}

\usepackage{newunicodechar}

\newunicodechar{≤}{\ensuremath{\leq}}
\begin{document}
\title{Dynamic In-context Learning with Conversational Models for Data Extraction and Materials Property Prediction}
\author{Chinedu E. Ekuma}
\altaffiliation{che218@lehigh.edu}
\affiliation{Department of Physics, Lehigh University, Bethlehem, PA 18015, USA}

\date{\today}

\begin{abstract}  
\noindent  
The advent of natural language processing and large language models (LLMs) has revolutionized the extraction of data from unstructured scholarly papers. However, ensuring data trustworthiness remains a significant challenge. In this paper, we introduce \texttt{PropertyExtractor}, an open-source tool that leverages advanced conversational LLMs like \texttt{Google gemini-pro} and \texttt{OpenAI gpt-4}, blends zero-shot with few-shot in-context learning, and employs engineered prompts for the dynamic refinement of structured information hierarchies – enabling autonomous, efficient, scalable, and accurate identification, extraction, and verification of material property data. Our tests on material data demonstrate precision and recall that exceed 95\% with an error rate of approximately 9\%, highlighting the effectiveness and versatility of the toolkit. Finally, databases for 2D material thicknesses, a critical parameter for device integration, and energy bandgap values are developed using \texttt{PropertyExtractor}. Specifically for the thickness database, the rapid evolution of the field has outpaced both experimental measurements and computational methods, creating a significant data gap. Our work addresses this gap and showcases the potential of \texttt{PropertyExtractor} as a reliable and efficient tool for the autonomous generation of various material property databases, advancing the field.
\end{abstract}    

\keywords{\texttt{PropertyExtractor}; large language models; autonomous data extraction; dynamic zero-shot-few-shot in-context learning; 2D materials property; thickness}

\maketitle 
\section{Significance Statement}
High-fidelity material property databases are essential for addressing various science and engineering challenges, including algorithm development and downstream tasks. Recent advancements in generative AI technology have significantly improved material property extraction from unstructured literature, but maintaining data accuracy and reliability remains a challenge. Here, we introduce a computational framework that addresses this issue by leveraging advanced large language models (LLMs), integrating zero-shot and few-shot in-context learning techniques for dynamic information hierarchy refinement. This facilitates autonomous, efficient, and scalable extraction and verification of material properties. The user-friendly architecture allows researchers to generate accurate databases with minimal LLM or programming expertise, thereby democratizing access.

\section{Introduction}
The advent of generative artificial intelligence, particularly through advances in large language models (LLMs) and natural language processing (NLP), represents a transformative shift in the ability to harness unstructured data across diverse fields, such as material science. The abundance of data, including journal articles, patents, and theses, could be overwhelming for manual data extraction, especially due to the rapid pace of publication and its voluminous nature. Efficiently extracting and utilizing this information for material characterization and discovery remains a significant challenge. Traditional methods for data extraction from unstructured texts in materials science have relied on labor-intensive setups, including the development of parsing rules and the identification of specific phrases, necessitating extensive model fine-tuning and re-training. Such processes are resource-consuming and time-intensive, often resulting in systems that are rigidly specialized for narrow tasks~\cite{KONONOVA2021102155,Gilligan2023}. Recent advancements in the \textit{generative pretrained transformer} series of LLMs, noted for their exceptional capabilities in generating coherent and contextually relevant text, summarizing content, and identifying relationships between concepts, offer a promising solution. These conversational LLMs automate the extraction of relevant information from extensive document collections, facilitating the creation of material property databases and accelerating the discovery of new materials~\cite{Polak2024,D3DD00113J,Dagdelen2024}.

Recent advances in the field of materials science have significantly benefited from the development of NLP techniques that aim to extract valuable information from unstructured research papers. These efforts have laid the foundation for sophisticated data analysis and extraction methods, allowing researchers to gain insights that were previously inaccessible~\cite{Polak2024,D3DD00113J,Gilligan2023,D3DD00019B}. Despite these advances, challenges remain in achieving high fidelity data and context-aware data extraction, particularly when dealing with complex physical properties and diverse scientific terminologies. Huang \& Cole fine-tuned a BERT model on battery-related publications to enhance a battery database, employing a ``question and answer'' (Q/A) strategy to extract specific device-level information. However, their method struggles with passages containing multiple device information and requires extensive pre-training on battery research papers before fine-tuning for the Q/A tasks~\cite{doi:10.1021/acs.jcim.2c00035}. More recently, Zheng \textit{et al.} developed a prompt-engineering approach, ChemPrompt with ChatGPT, focused on transforming text into tabular forms and summarizing scientific papers, leveraging the vast pretraining corpus to create semi-structured summaries~\cite{doi:10.1021/jacs.3c05819}. Similarly, studies by Castro and Pimentel explored ChatGPT's baseline chemistry knowledge and found that without advanced prompt engineering, the model's performance was suboptimal on straightforward chemistry tasks~\cite{doi:10.1021/acs.jcim.3c00285}. The work of Xie \textit{et al.} expanded the use of LLMs in materials science, fine-tuning them on a broad corpus for diverse tasks such as Q/A, inverse design, classification, and regression~\cite{Xie2023DARWIN}. Despite these advancements, these approaches often fail to extract structured representations of complex hierarchical entity relationships and do not generalize beyond the limitations of the pretraining corpus. Recent works also showed that the performance of LLM-based architectures in extracting material property information can be improved via engineered prompts~\cite{Dagdelen2024,Polak2024}. While these advances have significantly enhanced the structured information extraction of material property data, there is still an inherent issue of scalability and transferability, including challenges in extracting structured representations of complex hierarchical entity relationships generalizing outside of the pretraining corpus.

\begin{figure*}[htb!]
        \centering
       \includegraphics[width=\linewidth]{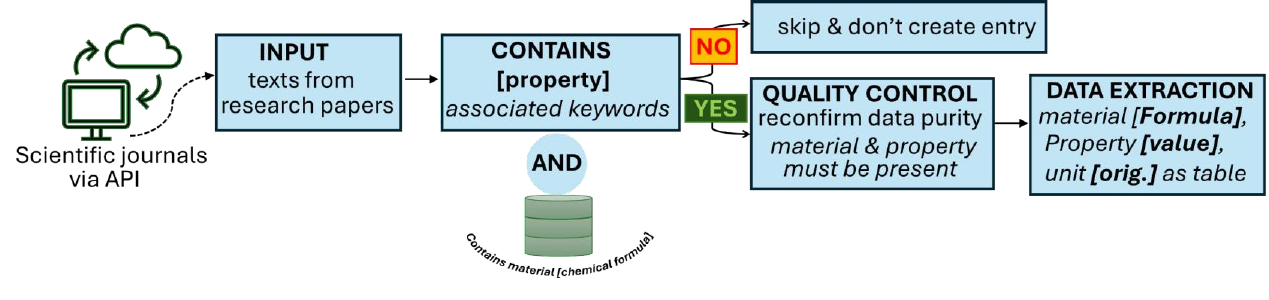}
        \caption{\textbf{Summarized flowchart of the \texttt{PropertyExtractor} code for obtaining structured dataset with a conversational large language model.} The flow diagram provides the basic ideas for each step of the process and illustrates the integration with the API for obtaining unstructured scientific papers. A more detailed flowchart is presented in Figure~\ref{fig1_2}.}
        \label{fig1_1}
\end{figure*}

In this paper, we introduce a \textit{blended dynamic zero-shot-few-shot in-context learning approach}. Here, task-specific instructions, such as ``Yes'' and ``No'' responses (zero-shot learning) are combined with non-prescriptive guidance (few-shot learning) that dynamically incorporates accurately performed tasks into the model, enhancing and providing a closed feedback loop for both scalability and predictability. In the zero-shot learning method, the model leverages its pre-existing (or ``general'') knowledge and understanding to generate responses or outputs relevant to tasks on which it was not specifically trained, based solely on the instructions given in the prompt. For instance, determining whether the sentence ``Graphene is a semiconductor'' is an instance of misconception could involve the following prompt:
\begin{verbatim}
Prompt:
Classify whether or not the text below
expresses graphene feature skepticism:

Text: "Graphene is a semimetal, not a
semiconductor, because it lacks a bandgap."
Classification:
\end{verbatim}
Recent works have shown that zero-shot learning applications using LLMs can yield reasonable results~\cite{Polak2024}. Nevertheless, zero-shot learning relies heavily on the model's ``general'' knowledge and may not perform optimally on specific downstream and domain-specific tasks, such as material property analysis and material science. Few-shot learning addresses these limitations by providing additional domain-specific examples to enhance the LLM’s understanding. The model then generalizes from these examples to perform the task effectively, even with minimal training data. Consequently, combining zero-shot and few-shot learning opens up new possibilities for employing LLMs in a broad range of applications without the need for extensive fine-tuning on specific tasks, making them more versatile, adaptable to various contexts inherent in material science, and scalable.

The concept of in-context learning (ICL) has emerged as a central paradigm for task adaptation in LLMs~\cite{Dong2023InContextLearning}, fundamentally enabling the model to adapt its behavior based on provided examples rather than undergoing resource-intensive fine-tuning of its internal parameters. ICL effectively leverages the ``context'' embedded within the model's prompt to adapt the LLM to specific downstream tasks, spanning a spectrum from zero-shot learning (where no additional examples are provided — task descriptive instructions) to few-shot learning (where several examples are offered). Leveraging the inherent semantic similarity, our approach dynamically selects contextually relevant outcomes from a set of accurately prior predicted tasks, thereby enhancing prompt engineering and scalability for the predictability of the large language model-based toolkit. Such an approach is pivotal in materials science, particularly for extracting critical property data such as those of novel two-dimensional (2D) materials~\cite{Novoselov2012,kastuar2022efficient,NAJMAEI2020110,Khanmohammadi2024}, which generally lack generalizable property databases compared to conventional bulk materials.

We implement this approach in the open-source code, \texttt{PropertyExtractor}, a Python-based computational framework designed to generate structured material property datapoint quadruplets of \textit{material, property value, original unit, method}. It employs a sophisticated combination of engineered prompts and dynamic zero-shot-few-shot in-context learning to identify data-rich sentences, whether these appear within tables or as single or multiple data values within the text. This approach not only extracts data but also verifies its accuracy, addressing common issues associated with factual inaccuracies typically observed in large language model responses. \texttt{PropertyExtractor} achieved significant improvements in model scalability, realizing a precision of approximately 96\%, a recall of 94\%, an accuracy of 90\%, an F1-score of 95\%, and an error rate of approximately 10\% on a constrained dataset of the thickness of 2D materials. For the extraction of energy bandgap values, \texttt{PropertyExtractor} demonstrated even better performance metrics. It achieved a precision of 96.81\%, a recall of 94.72\%, an F1-score of 95.21\%, an accuracy of 92.05\%, and an error rate of approximately 7.95\%. These results highlight the effectiveness and reliability of \texttt{PropertyExtractor} in accurately extracting critical material properties from scientific literature. These performance metrics surpass those reported in earlier studies~\cite{Polak2024,D3DD00113J,Gilligan2023}, highlighting its potential utility in both academic and industrial settings. Moreover, \texttt{PropertyExtractor} is designed as a software and leverages online LLM APIs, which allow users to train bespoke models without extensive knowledge of the internal workings of LLMs or Python skill. The user may simply treat the \texttt{PropertyExtractor} architecture as a black-box that transforms passages into precisely-formatted, structured summaries of material property data, thus enabling researchers with little NLP experience to use the toolkit. While \texttt{PropertyExtractor} is currently implemented for OpenAI gpt and Google gemini, its architecture is deliberately designed for easy adaptation to incorporate any conversational LLM. This flexibility is achieved by modularizing the component that interfaces with the LLM, allowing for simple modifications of a specific subroutine to accommodate alternative models. Such an adaptable system ensures that \texttt{PropertyExtractor} not only keeps pace with the rapid advancements in the field of LLMs but also leverages these improvements to enhance its functionality. The evolving landscape of LLMs development suggests a future where continual improvements in model performance could further amplify the efficacy of \texttt{PropertyExtractor}. Drawing parallels from the field of image generation, where prompt engineering has become a standard practice to secure high-quality results, a similar trend is expected in data extraction. Our approach, with its foundation in dynamical in-context learning, prompt engineering, and conversational prompts, positions \texttt{PropertyExtractor} as a versatile tool adaptable to the forthcoming generations of LLMs, ensuring high-quality data extraction across diverse applications.

\section{Results and Discussions}\label{backg}
\subsection{Data Extraction Workflow}
\begin{figure*}[htb]
        \centering
       \includegraphics[width=\linewidth]{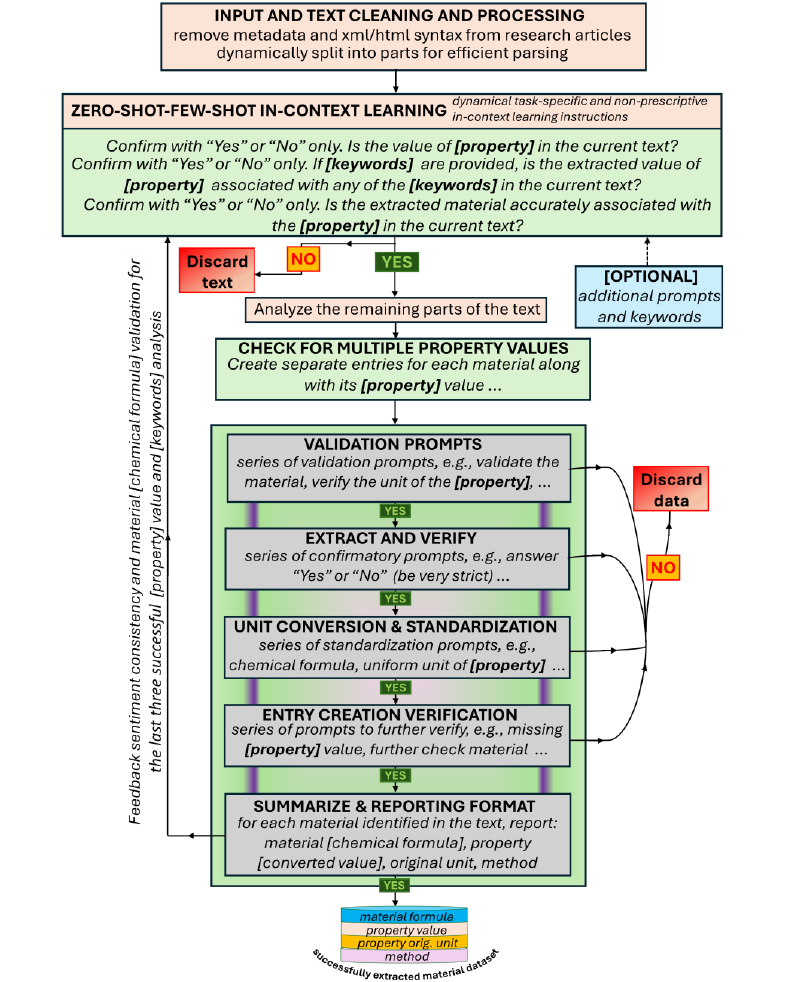}
        \caption{\textbf{Flowchart of the \texttt{PropertyExtractor} code for obtaining structured dataset with a conversational large language model.} The flow diagram is made up several interrelated prompt sections guiding the framework for the successful and accurate extraction of materials properties in the form of the quadriplets []\textit{material, property value, original unit, method}]. }
        \label{fig1_2}
\end{figure*}

Figure~\ref{fig1_1} shows a concise workflow of the integrative unstructured data collection via API and structured data extraction with the \texttt{PropertyExtractor}. In Figure~\ref{fig1_2}, we present a detailed workflow illustrating the essential processes and steps for extracting material properties from unstructured research papers. The data extraction process begins with data preparation. This stage involves programmatically gathering relevant scientific literature using various APIs and refining the content to create clean, unstructured data suitable for the conversational language model. The data curation is specifically designed to target texts mentioning the principal material property and may include an optional set of keywords related to the specific property, using an extensive set of keywords to cover the diverse terminology within the field. While the keywords are optional, they are essential for ensuring that the extracted data is relevant to the specified material property. For example, extracting energy bandgap specific to 2D materials could have ``band gap'' as the property and ``2D material'' as one of the keywords. The initial process of data extraction involves two distinct stages. The first stage is the data collection phase, where scientific papers are programmatically gathered using API. This is followed by the data preparation phase, where the collected papers are cleaned to remove XML/HTML tags and other syntactical clutter and then segmented into individual sentences for further processing. Following the cleaning of unstructured data, the next crucial step involves incorporating the LLM-based model, which is enhanced by dynamic zero-shot-few-shot in-context learning. This advancement refines the LLM’s conversational interactions with the data using contextually rich prompts. We found that integrating zero-shot and few-shot in-context learning with dynamic refinement through feedback on recently, accurately predicted outcomes enables the LLM to adaptively refine the data extraction process. This strategy significantly enhances extraction accuracy by minimizing errors and preventing data hallucinations. It effectively addresses the challenges posed by the diverse structures, complexity, and terminology found in unstructured scientific literature, ensuring meticulously calibrated responses for high-fidelity data extraction. The critical steps of this process are broken down as follows:

\begin{enumerate}
    \item \textbf{Initial data classification.} We begin with an initial screening using a simple relevancy prompt. This step is crucial for filtering out sentences that lack pertinent data about the property of interest.
    
    \item \textbf{Data extraction and validation.} This stage involves a sophisticated, multistage extraction process. Central to this phase is the application of dynamic zero-shot-few-shot in-context learning. The model not only learns from a carefully selected array of relevant examples but also incorporates feedback from three recent accurate predictions. This method enhances the model's understanding of complex scientific concepts, significantly improving the accuracy of data extraction. This approach involves the following (Figure~\ref{fig1_2}):
    \begin{itemize}
    \item \textbf{Employing engineered prompts.} The model utilizes engineered prompts that facilitate learning from prior examples. This foundational step improves the model's capability to parse and accurately extract data about material properties, setting the stage for more complex interactions.
    \item \textbf{Dynamically adjusting prompts.} As the model processes the data, it dynamically adjusts the prompts based on recent analyses. This adaptation captures the complexity and variability inherent in scientific texts, tailoring interactions to improve understanding and response accuracy.
    \item \textbf{Handling data variability.} The model is equipped with strategies to efficiently manage single- and multi-valued data points, enhancing its adaptability across various data extraction scenarios. For multi-valued scenarios, the model uses regex patterns and dynamic zero-shot-few-shot in-context learning to address the complexity of extracting and correctly associating each value with its respective material and property. This method begins with detecting the presence of multiple data points within a sentence, followed by tailored prompts that guide the model in differentiating and processing these values accurately. Additionally, the model employs prompts that encourage critical evaluation of its responses. This not only reduces errors but also significantly enhances the trustworthiness of the extracted data by ensuring that each data point is accurately captured and attributed.
    \item \textbf{Critique and uncertainty quantification.} The model can evaluate its own performance, critiquing its own outputs to explicitly acknowledge when essential data may be missing from the text. This precaution is designed to prevent the model from ``hallucinating'' or generating non-existent data to fulfill the task requirements. Additionally, to refine its accuracy further, the model employs uncertainty-inducing redundant prompts that encourage negative answers when appropriate. This strategy allows the model to critically reanalyze the text, avoiding the reinforcement of potentially incorrect previous answers. This iterative reevaluation process is crucial in minimizing errors and enhancing the trustworthiness of the extracted data.
    \item \textbf{Customization through user input.} Users can provide optional arguments for custom prompts and keywords tailored to specific properties being extracted, such as thickness in the case of 2D materials. Users can also assign weights to these keywords based on their priority and improve the significance of the keywords through synonyms, optimizing the model’s focus and relevance to the targeted extraction task.
    \item \textbf{Data extraction and standardization.} Enforce rigorous standardization and structured data entry to ensure uniformity of the property values, enhance consistency, and facilitate automated data processing.
\end{itemize}
\end{enumerate}

While the prompts associated with the data classification task are the first consequential prompts the model interacts, we have found that setting the stage by informing the model of the ``task description'' — explicitly telling the LLM that the task is to analyze scientific literature and identify certain properties — increases the predictability of the model. The data extraction process begins with a critical initial classification step, involving a preliminary prompt designed to assess the relevance of each sentence within a scholarly paper. The primary goal at this stage is to determine whether a sentence contains pertinent information regarding the material property of interest, such as the property (e.g., thickness), and the corresponding values and material names. Given the vast amount of content in scientific papers, even after narrowing the papers through initial keyword searches, the proportion of sentences containing relevant data is exceedingly low — typically around $13\%$. Hence, the immediate elimination of irrelevant sentences — those that do not contain at least the property and its value — is imperative to enhance efficiency and focus the analysis on potentially useful data. The LLM-based \texttt{PropertyExtractor} toolkit is designed to process input texts ranging from as small as an abstract to as large as the entire article, including tables. After initially classifying the entire text to determine the presence of relevant material properties, we have found that dynamically breaking the large texts into smaller chunks, such as paragraphs, improves efficiency and reduces contextual overload. LLMs often struggle with processing extremely long sequences of text due to computational limitations. Operational experience suggests that maintaining the brevity of the text passage is paramount for achieving the highest accuracy in data extraction. While it is technically feasible to expand the text passage further to capture more comprehensive data points, such expansions often do not justify the cost due to the marginal increase in accuracy and the potential detriment to the precision of the extraction process. Nevertheless, the methodology for text selection and the extent of text expansion can be fine-tuned based on the specific characteristics of different LLMs or the particular properties targeted in the analysis. Tailoring these parameters may lead to improvements in data extraction accuracy and efficiency in certain scenarios.

In the data extraction and validation phase following the initial data classification, we focus on extracting a quadruple of material features: chemical formula, property value, original unit, and method. While data are strictly created when the pair comprising the chemical formula of the material and the property value is present, we foresee broader applications of \texttt{PropertyExtractor}, such as discerning whether a material's property value is derived from computational methods or experimental techniques. This distinction is crucial for constructing a high-fidelity database. We employ a combination of task-descriptive instructions and non-prescriptive guidance (\textit{employing engineered prompts}) to facilitate dynamic in-context learning through a self-determined feedback mechanism (\textit{dynamically adjusting prompts}). Generic prompts often struggle to manage texts with multiple property values. We have found that utilizing regex patterns for straightforward cases and dynamic, context-sensitive prompting for more complex scenarios within the LLM architecture enables tailored data extraction processes that are both efficient and reliable for texts containing single or multiple values (\textit{handling data variability}). Self-assessment mechanisms are integrated at each step of the process (\textit{critique and uncertainty quantification}) to allow the model to evaluate its own performance and explicitly recognize instances where essential data may be missing. Through dynamic feedback, the model is equipped to prevent itself from reinforcing potentially inaccurate prior responses and to critically reanalyze the text. Iteratively reevaluating the extracted data is essential to improving reliability. The model critiques its own outputs to ensure no essential data is omitted and employs uncertainty-inducing redundant prompts that encourage negative answers when appropriate. This strategy prevents the model from ``hallucinating'' or fabricating data to meet task requirements, critically reanalyzing the text to avoid reinforcing incorrect previous answers. This iterative reevaluation process is crucial in minimizing errors and enhancing the trustworthiness of the extracted data.

Furthermore, it is important to recognize that despite the conversational model's capability to retain information throughout a dialogue, augmenting the dialogue with accurately predicted tasks is crucial for preserving detailed information about the overall nature and complexity of the type of text being analyzed. As discussions progress, the model may tend to overlook finer details. Thus, the conversational aspect and the strategy of information retention significantly enhance the quality of responses and underscore the benefits of using blended and structured prompts. The ability to retain crucial information in a conversational context is essential for ensuring both the accuracy and completeness of the extracted data. The extraction and standardization of data using LLMs present significant challenges. Although strict binary ``Yes'' or ``No'' responses have been proposed to streamline the generation of property entries,~\cite{Polak2024} this approach can restrict the precision achievable with LLMs. We have developed a more robust approach that combines the power of regular expressions (regex) with structured responses for enhanced property extraction. This nuanced strategy leverages the versatility of regex patterns alongside structured response frameworks, facilitating the extraction of relevant properties without the limitations imposed by binary choices. To ensure consistency and support downstream analysis, we first implement a rigorous standardization protocol. \textit{Unit harmonization:} Property values are converted to a standardized unit specified by the user (e.g., thickness in \AA), while original units are retained within the metadata for reference. This approach helps in reducing potential errors in data interpretation. \textit{Structured entry:} Each verified material entry is systematically compiled into a quadruple format: \textit{material} [chemical formula], \textit{property value} [converted value], \textit{unit} [original unit], and \textit{method} [method]. This structured format is efficiently extracted using regex patterns, which not only maintains the integrity and usability of the information but also enhances its utility for scientific analysis. By implementing such stringent protocols, we not only minimize uncertainty but also enhance the efficiency of data handling and analysis through streamlined automation.
\begin{figure*}[htb!]
        \centering
       \includegraphics[width=\linewidth]{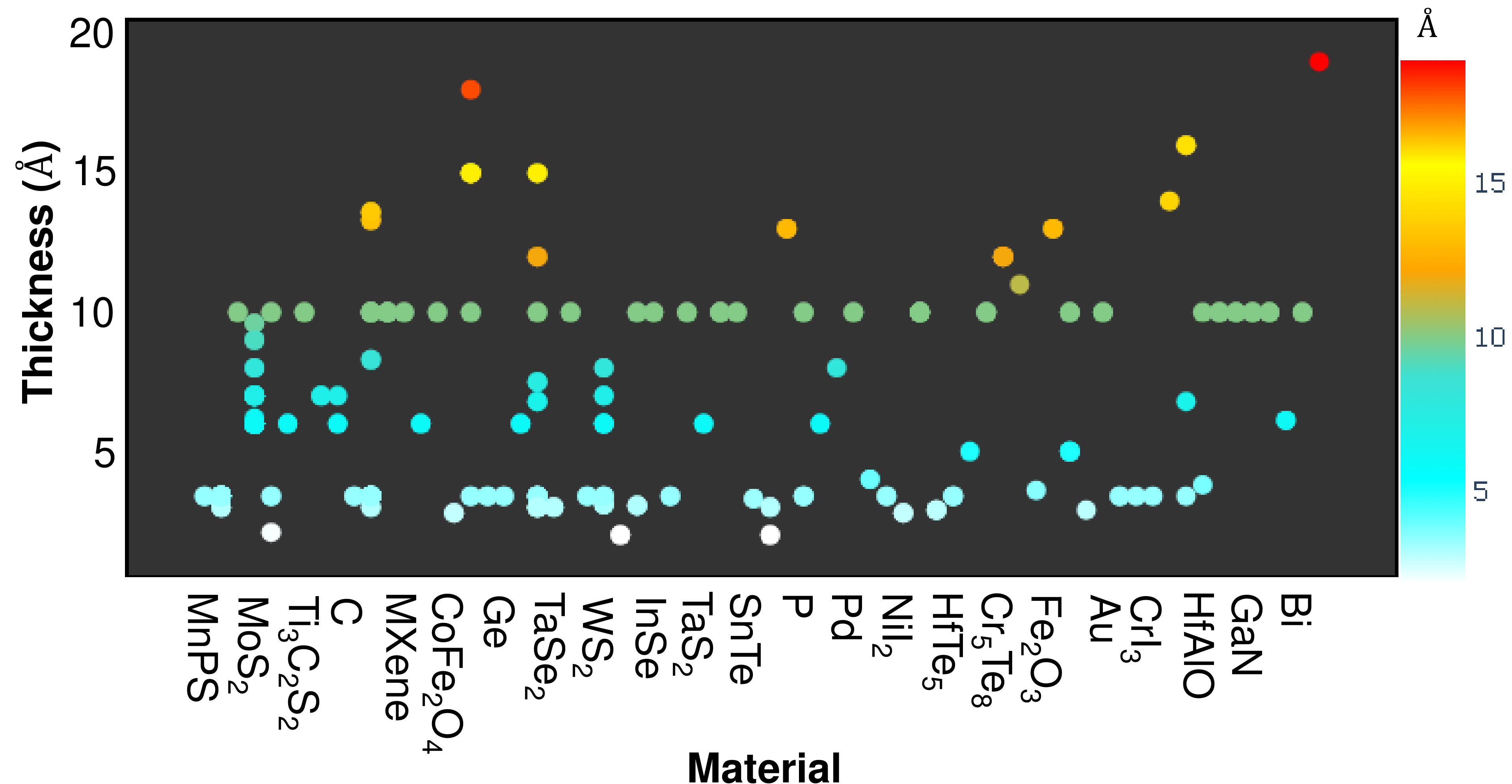}
        \caption{\textbf{Material property database with \texttt{PropertyExtractor}.} A snapshot of the extracted thickness data for atomically thin 2D materials, illustrating the range and diversity of the autonomously obtained database.}
        \label{fig3}
\end{figure*}

\subsection{Performance Evaluation}
In this study, we provide two materials property databases: the thickness of 2D-based materials and the energy bandgap of materials. The database for thickness has been post-processed to remove entries that do not belong to 2D materials. Accounting for the exact thickness, such as being a single-layer, is not feasible since this parameter is not reported in most literature. For the energy bandgap, we provide the raw output obtained using the \texttt{PropertyExtractor}. We evaluated the performance of \texttt{PropertyExtractor} by developing databases for the energy bandgap of materials and the thickness of 2D-based materials. The choice of these databases was intentional to address two critical aspects: testing the scalability of \texttt{PropertyExtractor} in obtaining hard-to-get material property features like the thickness of 2D-based materials and validating its performance on a well-established material property, the energy bandgap, which is readily available in many databases.

The focus on thickness is driven by two main factors. Despite its crucial importance for integrating 2D materials into devices, there is currently no database or computational framework for determining the thickness of these materials. The thickness of 2D materials significantly influences their electronic structure, optical properties, mechanical behavior, and chemical reactivity. For example, MoS$_2$ exhibits a transition from an indirect to a direct bandgap when reduced to a single layer, enhancing its suitability for optoelectronic devices. Similarly, the mechanical properties of graphene and its surface-to-volume ratio, which are critical for catalysis and sensor applications, are thickness-dependent. This property is also vital for the performance of energy storage devices, impacting ion transport and storage capacity. Additionally, the thickness dataset poses a significant challenge due to the potential for confusion with similar properties, such as layer count or interlayer spacing, which require precise differentiation. Hence, it provides a robust testbed to evaluate the scalability and capability of \texttt{PropertyExtractor}, demonstrating its effectiveness in accurately discerning and extracting complex data amidst closely related variables.

\noindent \textit{Text Mining and Database Generation}.
The acquisition of thickness information for 2D materials is challenging due to data scarcity. Despite having access to APIs, obtaining reliable thickness values from a single source remains difficult. We utilized several APIs, including Elsevier's ScienceDirect API, CrossRef REST API, and PubMed API. API calls were conducted using a combination of the keywords ``Thickness of 2D materials'' and variations such as ``Thickness measurement of 2D materials,'' ``Thickness determination of two-dimensional materials,'' or ``Characterization of 2D material thickness.'' These texts were subsequently pre-processed and cleaned to retain only those containing abstracts or full texts and the keyword ``thickness,'' resulting in a collection of 458 full-text articles from the Elsevier's ScienceDirect API, 424 abstracts from the PubMed API, and 8,387 abstracts from the CrossRef REST API.

Fulltexts or abstracts were processed using the Google gemini-pro API conversational language model within Python 3.10.12, employing our dynamic zero-shot-few-shot in-context learning and prompt engineering approach as implemented in \texttt{PropertyExtractor}. This process yielded 1,221 raw data points, which, after cleaning, resulted in 1,015 data points. Further post-processing standardized these to 584 data points with units specified in \AA. Among these, some records pertained to materials that might not be fully atomically thin 2D materials, despite initial keyword searches designed to exclude such instances. The inclusion of non-2D materials does not reflect a deficiency in the capabilities of our conversational LLM, as these data were present in the literature sourced. Restricting the database exclusively to atomically thin 2D materials yielded a final standardized database containing thickness values for about 43\% of the dataset. The extracted thickness dataset is dominated by the well-known 2D materials such as graphene, transition metal dichalcogenides, BN, phospherene, etc.(Figure~\ref{fig3}). Within this dataset, 68 thickness values corresponded to unique materials. It is important to note that in cases where texts explicitly mention 2D material thickness, the model extracted these instances with nearly 100\% accuracy, demonstrating the robustness of \texttt{PropertyExtractor} in handling texts with less ambiguous keyword descriptions.

To develop the energy bandgap database using \texttt{PropertyExtractor}, we applied a similar procedure as we did for the thickness of 2D-based materials. The only difference was that the primary and secondary keywords were replaced with ``bandgap'' and ``band gap,'' respectively. To obtain the source research articles, we performed a search query exclusively employed the PubMed API, which initially provided us with 12,102 independent entries. After processing the unstructured data from the PubMed API, we refined these entries down to 9,987 unique full texts or abstracts following further pre-processing steps. Utilizing the \texttt{PropertyExtractor} in conjunction with the Google Gemini-pro API conversational language model within Python 3.10.12, we then processed the 9,987 unique entries. This comprehensive processing resulted in 1,238 entries containing energy bandgap values and identified 561 unique materials.

\noindent \textit{Evaluation Methodology for \texttt{PropertyExtractor}.} 
Given the lack of a ground truth database for the thickness of 2D materials, we developed a unique approach to evaluate the performance of \texttt{PropertyExtractor}. We curated a dataset from the scientific literature, extracting data to establish a ground truth. From approximately 12,000 fulltexts or abstracts, we selected 100 relevant ones containing thickness data. This selection process was targeted to ensure that the model only processes potentially useful data, optimizing computational resources and focusing on relevant texts. We defined our ground truth as 50 manually extracted data points from these texts, focusing on triplets comprising the material [chemical formula], its thickness value, and the measurement unit. This methodology allows for a precise evaluation by testing the model against a well-defined set of data, which is crucial for assessing the tool's capability to accurately extract specific material properties from complex scientific texts. Similar approach is used to establish the ground truth data specific to the energy bandgap. The validation process is carried out using both \textit{gemini-pro} and \textit{gpt-4}.

\begin{figure}[htb!]
        \centering
       \includegraphics[width=\linewidth]{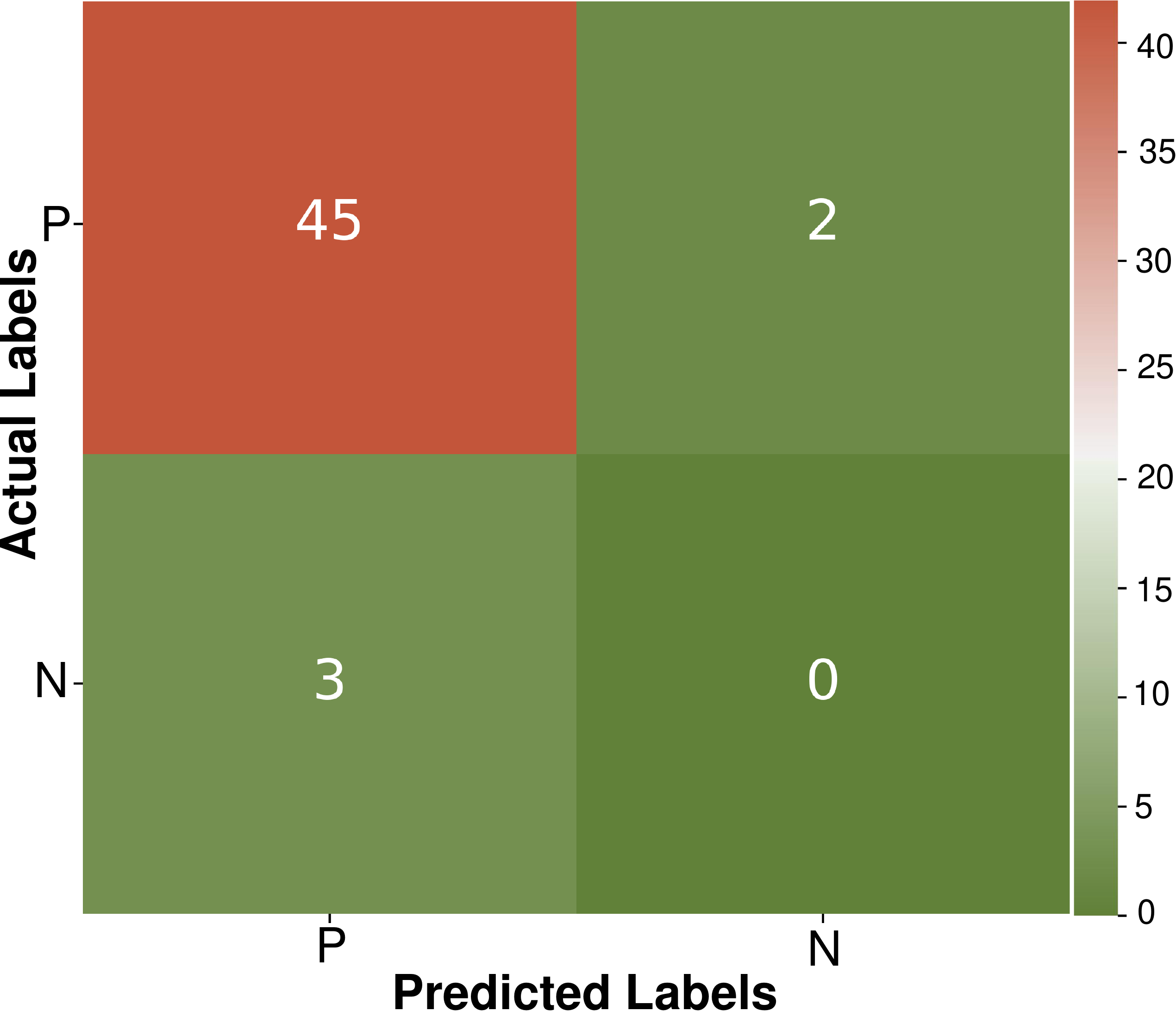}
        \caption{\textbf{Performance Evaluation of \texttt{PropertyExtractor}.} Confusion matrix comparing the ground truth with data extracted by \texttt{PropertyExtractor}, showcasing 45 true positives, 2 false positives, and 3 false negatives, which are used to calculate the model's precision, recall, accuracy, and error metrics.}
        \label{fig4}
\end{figure}

To design the statistical metrics for evaluating \texttt{PropertyExtractor}, we generate a confusion matrix (Figure~\ref{fig4}. This matrix compares predicted values to actual target values, providing counts of true positives (\textit{TP}), true negatives (\textit{TN}), false positives (\textit{FP}), and false negatives (\textit{FN}), thereby facilitating a detailed analysis of the model's accuracy, precision, recall, and other metrics. The evaluation process is outlined as follows: (1) \textit{Ground Truth and Extracted Data:} We define ground truth as the hand-extracted triplets of material, value, and unit from the text, with the model producing zero or more extracted triplets. (2) \textit{Evaluation Criteria:} \textit{FP} arise when extracted triplets are present but not in the ground truth; \textit{FN} occurs when the ground truth contains triplets but the extracted data does not; \textit{TP} are noted when each extracted triplet matches a ground truth triplet in terms of material, value, and unit. Since our focus is on correct/incorrect labels, \textit{TN} is not applicable, implying $\textit{TN} \approx 0$. (3) \textit{Comparison of Triplets:} A \textit{TP} is counted if a matching triplet is found. A \textit{FP} is counted if a non-matching triplet is found. Multiple extracted triplets compared to a single ground truth triplet count as \textit{TP} for the first correct match, with subsequent matches counted as \textit{FP}. For multiple ground truth triplets vs. extracted data, the absence of extracted triplets results in multiple \textit{FN}, and any match with ground truth counts as a \textit{TP}. (4) \textit{Definition of Equivalence:} Extracted triplets must exactly match ground truth in units and values, and the material names must uniquely identify the same material system.

\noindent \textit{Evaluation and Accuracy Analysis}.
The primary statistical metrics employed to evaluate the performance of \texttt{PropertyExtractor} are precision (P), recall (R), F1-score (F1$_s$), accuracy (Acc), and the error rate (E$_r$) given as:
\begin{align*}
    \text{P} & = \frac{TP}{TP + FP};\,  \text{R} = \frac{TP}{TP + FN}; \, \text{F1}_s  = 2 \times \frac{P \times R}{P + R}  \\
    \text{Acc} & = \frac{TP}{TP + FP + FN}, \,  \text{E}_r  = \frac{FP + FN}{TP + FP + FN}.
\end{align*}
We first evaluated the performance of \texttt{PropertyExtractor} in developing a database for the thickness of 2D materials. Utilizing \texttt{PropertyExtractor} with \textit{gemini-pro}, the achieved metrics are as follows: precision of 95.74\%, recall of 93.75\%, F1-score of 94.73\%, accuracy of 90.00\%, and an error rate of approximately 10.00\%. These values indicate that \texttt{PropertyExtractor} correctly predicts a triplet 95.56\% of the time (precision) and successfully identifies 93.48\% of all relevant triplets (recall). The accuracy of 89.58\% reflects the overall correctness of the tool, accounting for both positives and negatives. An F1-score of approximately 94.51\% suggests a balanced model in terms of precision and recall. The error rate, at around 10.42\%, demonstrates the model's effectiveness, with a small proportion of the predictions being incorrect. Furthermore, when evaluations are specifically based on the material and standardized thickness values, the performance notably improves, exemplified by a precision of approximately 98\%. We note that the accuracy using \textit{GPT-4} is similar to that obtained with \textit{gemini-pro}, but \textit{GPT-4} performs better in the effective determination of the method used and in discerning data that are provided as a range.

The performance metrics for extracting the energy bandgap values are slightly improved compared to the thickness data, which we attribute mainly to the well-defined nature of energy bandgap in scientific articles. The achieved metrics are precision of 96.81\%, recall of 94.72\%, F1-score of 95.21\%, accuracy of 92.05\%, and an error rate of approximately 7.95\%. While a large automatically extracted database of bandgap values for materials has been developed previously,\cite{Dong2022,Kim2020} direct quantitative comparison is not straightforward. However, the histogram of values obtained from the previous database exhibits a very similar shape in terms of material types to the data obtained here, further supporting the robustness of our data. Moreover, many well-known semiconductors and their bandgap values are adequately captured in our database. It is important to note that the database of energy bandgap developed here is more diverse and heterogeneous, having been extracted from various first-principles calculations and diverse experimental techniques. This diversity in data sources underlines the robustness of \texttt{PropertyExtractor} in handling a wide range of information and providing reliable outputs.

\section{Conclusions}~\label{conclusion}
This paper demonstrates the power of zero-shot and few-shot in-context dynamical learning when integrated within conversational LLMs such as \textit{Google gemini-pro} and \textit{OpenAI GPT-4} as implemented in the open-source toolkit, \texttt{PropertyExtractor}. \texttt{PropertyExtractor} provides an easy-to-use computational architecture to achieve accurate, scalable, and transferable material property data extraction, enabling the creation of high-fidelity databases for use with downstream applications and models such as machine learning applications and the construction of knowledge graphs. We demonstrate the effectiveness of \texttt{PropertyExtractor} by generating unique databases for the thickness of 2D materials and the energy bandgap of materials, achieving impressive accuracy metrics. For the thickness of 2D materials, we achieved precision above 95\%, recall of approximately 93\%, accuracy of approximately 90\%, F1-score of approximately 95\%, and an error rate of only approximately 10\%. This database addresses a critical gap in materials science for 2D-based materials, where thickness information is crucial to harnessing the intrinsic thickness-dependent features of 2D materials. For the energy bandgap, the performance metrics are even more impressive, with a precision of 96.81\%, recall of 94.72\%, F1-score of 95.21\%, accuracy of 92.05\%, and an error rate of approximately 7.95\%. The high precision and recall metrics indicate that \texttt{PropertyExtractor} is highly effective in accurately identifying and extracting energy bandgap values from scientific literature. The adaptability of \texttt{PropertyExtractor} and its model independence ensure it can evolve alongside advancements in LLM architecture. As newer and more advanced LLMs come into existence, the efficacy and applicability of \texttt{PropertyExtractor} are poised to broaden, providing unprecedented opportunities for sophisticated data extraction across diverse scientific fields. The integration of conversational language models within \texttt{PropertyExtractor} showcases the potential of zero-shot and few-shot learning in dynamically understanding and processing complex scientific information, thereby enhancing the way material properties are extracted and utilized.

\section{Method}
\texttt{PropertyExtractor} is a conversational large language model toolkit designed for extracting physical properties from scientific articles. It incorporates a dynamic zero-shot-few-shot in-context learning architecture. Although \texttt{PropertyExtractor} is highly adaptable and can utilize various LLM models, it is currently optimized for use with Google Gemini Pro and OpenAI GPT-4. \texttt{PropertyExtractor} is designed to leverage the latest Python libraries; specifically, we used Python 3.10.12.

\textit{Installation.} The preferred installation method is via \texttt{pip}, the Python package manager, which simplifies the installation process. To install, execute the command: \texttt{pip install -U propertyextract}. Alternative traditional methods, such as downloading the sources and manual installation, are also available. Once installed, the essential input files required for setting up property extraction from unstructured scientific papers can be generated by running the command \texttt{propertyextract -0}. This command produces the main \texttt{PropertyExtractor} input file, \texttt{extract.in}, as described in Table~\ref{parameters_propertyextractor}, along with two optional files: \texttt{keywords.json} and \texttt{additionalprompt.txt}. These files offer further customization options for both the keywords specific to the property being extracted and the additional custom prompts, respectively. The essential parameters necessary for successfully running the model configuration used in this paper are presented in Table~\ref{parameters_propertyextractor} and detailed below.
\begin{itemize}[noitemsep, nolistsep]
    \item \texttt{model\_type}: Type of the model used (gemini/chatgpt).
    \item \texttt{model\_name}: Specific name of the model (gemini-pro/gpt-4).
    \item \texttt{property\_name}: The physical property being extracted (e.g., thickness).
    \item \texttt{property\_unit}: Harmonized unit for the physical property (e.g., angstrom, eV).
    \item \texttt{temperature}: The temperature condition under which the model operates, in degrees Celsius (default is ``0.0'').
    \item \texttt{top\_p}: The probability threshold for the model's output generation (e.g., 0.95).
    \item \texttt{max\_output\_tokens}: Maximum number of output tokens generated by the model (e.g., 80).
    \item \texttt{use\_keywords}: Toggle to determine whether keywords should be utilized in processing (True/False).
    \item \texttt{additional\_prompts}: File containing additional prompts that can be used to guide the model (e.g., additionalprompt.txt).
    \item \texttt{inputfile\_name}: Name of the CSV file containing the input data.
    \item \texttt{column\_name}: The specific column in the input file to be used for data extraction (e.g., Text).
    \item \texttt{outputfile\_name}: Name of the file where the processed output data will be saved (e.g., output.csv).
\end{itemize}

\begin{table}[htb]
    \centering{
        \caption{The control parameters and the representative values used in the \texttt{PropertyExtractor} code. \label{parameters_propertyextractor} }
    \begin{tabular}{cc}
        \hline
        \hline
        Parameters & Values \\
        \hline
        \texttt{model\_type} & \texttt{gemini/chatgpt} \\
        \texttt{model\_name} & \texttt{gemini-pro/gpt-4} \\
        \texttt{property\_name} & \texttt{thickness} \\
        \texttt{property\_unit} & \texttt{angstrom} \\
        \texttt{temperature} & \texttt{0.0} \\
        \texttt{top\_p} & \texttt{0.95} \\
        \texttt{max\_output\_tokens} & \texttt{80} \\
        \texttt{use\_keywords} & \texttt{True} \\
        \texttt{additional\_prompts} & \texttt{additionalprompt.txt} \\
        \texttt{inputfile\_name} & \texttt{thickness.csv} \\
        \texttt{column\_name} & \texttt{Thickness} \\
        \texttt{outputfile\_name} & \texttt{processed.csv} \\
        \hline
        \hline
    \end{tabular}
}
\end{table}

\section{Acknowledgments}
This work was supported by the U.S. Department of Energy, Office of Science,
Basic Energy Sciences under Award DE-SC0024099. Computational support is provided by the Computational materials research group at Lehigh University.

\section*{Data Availability}
The extracted database of thickness of 2D-based materials is available at\href{https://github.com/gmp007/PropertyExtractor}{PropertyExtractor@github} and via Zenodo at \href{https://doi.org/10.5281/zenodo.11205500}{10.5281/zenodo.11205500}. The thickness database developed as part of this software is included in the code website.

\section*{Code Availability}
The code, \texttt{PropertyExtractor}, developed in this paper is available at GitHub: \href{https://github.com/gmp007/PropertyExtractor}{PropertyExtractor@github} and via Zenodo at \href{https://doi.org/10.5281/zenodo.11205500}{10.5281/zenodo.11205500}. The code repository includes an easy-to-follow readme and a folder containing the data used in validating the performance. Additionally, the repository provides detailed instructions on setting up and using \texttt{PropertyExtractor}, ensuring that users can replicate the results presented in this paper and apply the tool to their own datasets.

\end{document}